\documentclass[aps,prb,preprint,groupedaddress]{revtex4-1}

\usepackage{graphicx}
\usepackage{bm}
\usepackage{txfonts}
\usepackage{color}
\graphicspath{ {figures/}}
\begin{document}


\title{Possible occurrence of superconductivity by the $\pi$-flux Dirac string formation due to spin-twisting itinerant motion of electrons}


\author{Hiroyasu Koizumi }
\affiliation{Division of Quantum Condensed Matter Physics, Center for Computational Sciences, University of Tsukuba, Tsukuba, Ibaraki 305-8577, Japan}


\date{\today}

\begin{abstract}
We show that the Rashba spin-orbit interaction causes spin-twisting itinerant motion of electrons in metals and realizes the quantized cyclotron motion of conduction electrons without an external magnetic field.  From the view point of the Berry connection, the cause of this {quantized}  motion is the appearance of a non-trivial Berry connection ${\bf A}^{\rm fic}=-{\hbar \over {2e}}\nabla \chi$ ($\chi$ is an angular variable with period $2\pi$)
that generates $\pi$ flux (in the units of $\hbar=1, e=1,c=1$) inside the nodal singularities of the wave function (a ``Dirac string'')
along the centers of spin-twisting.

Since it has been shown in our previous work\cite{koizumi2019} that the collective mode of $\nabla \chi$ is stabilized by the electron-pairing and generates supercurrent, the $\pi$-flux Dirac string created by the spin-twisting itinerant motion will be stabilized by the electron-pairing and produce supercurrent.
\end{abstract}


\maketitle

\section{Introduction}

Dirac considered a string of singularities of a wave function with flux through it; he showed that a magnetic monopole should exists at 
a terminal end of the string \cite{Monopole}. The vortex line in a type II superconductor may be considered as a realization of such an object with a magnetic monopole at a terminal end of it at the surface of the superconductor.

The vortex in the superconductor is explained by the emergence of a vector potential 
  \begin{eqnarray}
{\bf A}^{\rm fic}=-{\hbar \over {2e}} \nabla \chi
\label{Afic}
\end{eqnarray}
which accompanies the electromagnetic vector potential ${\bf A}^{\rm em}$, 
where $\hbar$ is Planck's constant divided by $2\pi$, $-e$ is the electron charge, and $\chi$ is an angular variable with period $2\pi$.
The sum of the electromagnetic vector potential ${\bf A}^{\rm em}$ and ${\bf A}^{\rm fic}$ 
 \begin{eqnarray}
{\bf A}^{\rm eff}={\bf A}^{\rm em}+{\bf A}^{\rm fic}
\end{eqnarray}
is an effective gauge invariant vector potential existing in superconductors \cite{Weinberg}. 

The standard theory of superconductivity is the BCS theory \cite{BCS1957}. It was originally developed from the energy gap model of Bardeen \cite{Bardeen1955}, and identified the cause of the energy gap as the electron pair formation. 
The BCS theory has been successfully predicted the superconducting transition temperature T$_c$, where T$_c$ is given as the energy gap formation temperature. The appearance of $\chi$ in ${\bf A}^{\rm fic}$ is due to the use of the particle number non-fixed wave function in the BCS theory; namely, it is attributed to the $U(1)$ gauge symmetry breaking, thus, it has been believed that the particle number non-fixed formalism is crucial for a superconductivity theory \cite{Anderson66,Anderson}.

Due to the success of the BCS theory, many researchers had thought that superconductivity was a completely solved problem; however, the high temperature superconductivity found in cuprates\cite{Muller1986} has proved it is not so. The superconductivity in the cuprate (the {\em cuprate superconductivity}) is markedly different from the superconductivity explained by the BCS theory (the {\em BCS superconductivity}). Apart from the very high superconducting transition temperature, differences include,

\begin{enumerate}
\item[1)] The normal state from which the superconducting state emerges is a doped Mott-insulator \cite{AndersonBook} although the BCS superconductivity assumes the band metal for the normal state.

\item[2)] Local magnetic correlations in the superconducting state and a close relation between the magnetism and superconductivity have bee observed in the cuprate \cite{Neutron,MagneticRIXS}, while the magnetism is harmful for the BCS superconductivity.

\item[3)] The superconducting coherence length of the cuprate is in the order of lattice constant (nano-scale)\cite{Sebastian2019}, while it is assumed to be much larger than the lattice constant in the BCS superconductivity.

\item[4)] The superconducting transition temperature for the optimally doped cuprate is given by the stabilization temperature of the nano-sized loop currents \cite{Kivelson95,HKoizumi2015B,Koizumi2017}, while it is given by the energy gap formation temperature in the BCS superconductivity.

\item[5)] The hole-lattice interaction is very strong and small polarons and bi-polarons are created in the cuprate \cite{Bianconi,Muller2007b,Miyaki2008,oyanagi}, while the BCS superconductivity does not assume such a strong electron-lattice interaction that forms small polarons.
 \end{enumerate}
In spite of more than 30 years of research, no widely-accepted theory exists for the mechanim of the cuprate superconductivity.  It is very plausible that a drastic departure from the BCS theory is needed for the elucidation of the cuprate superconductivity. 

In order to explain the cuprate superconductivity, a new supercurrent generation mechanism where ${\bf A}^{\rm fic}=-{\frac \hbar {2e}} \nabla \chi$ appears as the Berry connection \cite{Berry,BMKNZ} has been put forward
\cite{Koizumi2011,HKoizumi2013,HKoizumi2014,HKoizumi2015}. In this theory, a macroscopic supercurrent is generated as a collection of {\em spin-vortex-induced loop currents} (SVILCs), where the SVILC is a superconducting-coherence-length-sized loop current induced by a spin-vortex (SV) created around each doped hole in the CuO$_2$ plane.
It explains a number of experimental results in the cuprate superconductors \cite{Hidekata2011};
\begin{enumerate}

\item[1)] Nonzero Kerr rotation in zero-magnetic field after exposed in a strong magnetic field \cite{Kerr1}.

\item[2)] The change of the sign of the Hall coefficient with temperature change \cite{Taillerfer07b}.

\item[3)] The suppression of superconductivity in the $x=1/8$ static-stripe ordered sample \cite{Neutron}.

\item[4)] A large anomalous Nernst signal, including its sign-change with temperature change \cite{Nernst2005}.

\item[5)] The hourglass-shaped magnetic excitation spectrum \cite{Neutron}.

\item[6)] Fermi-arc observed in the AEPES \cite{ArpesRev}.

\end{enumerate}

Actually, the new supercurrent generation mechanism does not require the electron-pair formation for the supercurrent generation (this does not mean that the electron pairing is not relevant to the cuprate superconductivity); however, the resulting supercurrent explains the flux quantum $\Phi_0=h/2e$ and the voltage quantum $V_0=hf/2e$ ($f$ is the frequency of the radiation field).

At this point I would like to point out a serious misfit that was found in the ac Josephson effect \cite{Koizumi2011,HKoizumi2015}; it was noticed that the boundary condition employed for the standard derivation of the ac Josephson effect and that in the real experimental situation were different, and that charge $q=-e$ should be used for the charge on the particle tunneling through the Josephson junction instead for $q=-2e$ used by Josephson \cite{Josephson62}. Since the Berry connection origin explains the observed ac Josephson effect with  $q=-e$, 
it is suggested that the Berry connection origin of ${\bf A}^{\rm fic}$ may be more in accordance with the experiment than the $U(1)$ gauge symmetry breaking origin. Note that  the $q=-e$ electron transfer is possible if the two superconductors in the junction is in such a close contact that the Bogolibov quasiparticle excitations are absent during the electron transfer between them and it is accompanied by simultaneous transferring of electrons between the superconductors and the leads connected to them \cite{Koizumi2011,HKoizumi2015,koizumi2019}.

Motivated by the above developments, we have reinvestigated the superfluidity problem in general \cite{koizumi2019}.
Then, we have found that the particle number non-fixed formalism, such as the standard BCS formalism and the Bogoliubov-de Gennes formalism \cite{deGennes}, can be cast in a particle number fixed formalism if the Berry connection is employed. In other words,
the Berry connection put the Bogoliubov transformation into the particle number fixed form, and replaces the phase variable arising from the Bogoliubov transformation by the Berry phase from the Berry connection. In this way, the $U(1)$ gauge symmetry breaking origin of ${\bf A}^{\rm fic}$ may be replaced by the Berry connection origin.

Since the persistent current in topological insulators can be attributed to the Berry connection \cite{Kane2005}, the Berry connection may be the unified ingredient for persistent current generation in superconductors and topological insulators.

 In the present work, we put forward a possible appearance of ${\bf A}^{\rm fic}$ in the BCS superconductor from the view point of the Berry connection origin. In this mechanism we add a very small interaction term, the spin-orbit interaction
\begin{eqnarray}
{ e \over {2m^2c^2} }{\bf s}\cdot[{\bf E}^{\rm em} \times ({\bf p}+e{\bf A}^{\rm em})]
\end{eqnarray}
in the Hamiltonian \cite{Dirac},
where ${\bf s}$ is the electron spin angular momentum, $m$ is electron mass, and ${\bf E}^{\rm em}$ is an electric field.
When this interaction affects conduction electrons, it is called the Rashba spin-orbit interaction
\cite{Rashba}. Since the internal electric field ${\bf E}^{\rm em}$ exists more or less in any materials, the Rashba interaction exists more or less in any materials. We consider the case where the Rashba interaction energy is much smaller than the energy gap created by the electron-pairing in this work. 

The organization of the present work is following: in Section II, the quantized motion of  Bloch electrons under the influence of a magnetic field and the Rashba spin-orbit interaction is investigated. It is shown that the quantized cyclotron motion occurs even without an external magnetic field due to the existence of  ${\bf A}^{\rm fic}$ arising from the spin-twisting itinerant motion of electrons.
In Section III, the energy gap for the BCS model is obtained for the case where the electron pairing occurs between $({\bf k}, {\bf s}_0({\bf r}))$ and $(-{\bf k}, -{\bf s}_0({\bf r}))$, where ${\bf k}$ is the wave vector and ${\bf s}_0({\bf r})$ is the spin for the electron that depends on the coordinate ${\bf r}$; this coordinate dependence of the spin arises from the spin-twisting itinerant motion of electrons.
In Section IV, the reduction of the kinetic energy due to the spin-twisting itinerant motion is investigated. The energy reduction is shown to be optimum when the Berry connection is given by ${\bf A}^{\rm fic}=-{\hbar \over {2e}} \nabla \chi$, and the Meissner effetc occurs due to the existence of ${\bf A}^{\rm fic}$.
In Section V, the Berry connection from the many-body wave function ${\bf A}^{\rm MB}$ previously introduced \cite{koizumi2019} is shown to be identified as ${\bf A}^{\rm fic}$. This identification is important in relation to our previous work \cite{koizumi2019} since it is shown there that ${\bf A}^{\rm MB}$ is stabilized by the electron-pairing interaction, giving rise to non-trivial ${\bf A}^{\rm fic}$ for superconductivity.
Section VI is the section for concluding remarks, where we succinctly summarize part of our previous work \cite{koizumi2019} by presenting the particle-number fixed version of the BCS ground state. It is argued that the formalism using the particle number changing operators given in our previous work\cite{koizumi2019} and the BCS formalism have the same mathematical structure, thus, the both yield the same results except the origin of the ac Josephson effect.
 
 \section{Appearance of spin-twisting itinerant motion of Bloch electrons under the influence of the Rashba spin-orbit interaction}
 \label{section5}

In this section we consider the quantized motion of Bloch electrons under the influence of the Rashba interaction and magnetic field.
We use the method of the periodic-orbit quantization for this purpose. It is known that the quantized energy obtained in this way gives an accurate energy\cite{Gutzwiller}.

In order to obtain the periodic orbit, we use the wave-packet dynamics formalism \cite{Niu}. In this formalism, the motion of the center of the wave packet corresponds to the classical motion of the electron. The force for the classical motion can be evaluated using the wave packet localized both in the real coordinate space ${\bf r }$ and the wave vector space ${\bf q}$ under the constraint of the Heisenberg uncertainty condition \cite{DiracWP}.

Let us consider electrons in a single band and denote its Bloch wave as
\begin{eqnarray}
|\psi_{\bf q}\rangle = e^{i {\bf q} \cdot {\bf r}} |u_{\bf q}\rangle
\end{eqnarray}
where ${\bf q}$ is the wave vector and $|u_{\bf q}\rangle$ is the periodic part of the Bloch wave.

It satisfies the Schr\"{o}dinger equation,
\begin{eqnarray}
H_0[{\bf q}]|u_{\bf q}\rangle ={\cal E}({\bf q}) |u_{\bf q}\rangle,
\end{eqnarray}
where $H_0$ is the zeroth order single-particle Hamiltonian for an electron in a  periodic potential.

According to the wave packet dynamics formalism, $H_0[{\bf q}]$ is modified as
\begin{eqnarray}
H_0[{\bf q}] \rightarrow H_0 \left[{\bf q}+{ e \over {\hbar}} {\bf A}^{\rm em}({\bf r})\right].
\end{eqnarray}
in the presence of the magnetic field ${\bf B}^{\rm em}=\nabla \times {\bf A}^{\rm em}$.

Using the Bloch waves, a wave-packet centered at coordinate ${\bf r}_c$ and with central wave vector ${\bf q}_c$ is constructed as
\begin{eqnarray}
\langle {\bf r}|({\bf q}_c, {\bf r}_c) \rangle &=& \int d^3q \ a({\bf q},t) 
\langle {\bf r}|\psi_{ {\bf q}}\rangle \Sigma_1({\bf r})
 \label{wp}
\end{eqnarray}
where $a({\bf q})$ is a distribution function, and $\Sigma_1({\bf r})$ is a spin function.

An important point is that we use the spin function that depends on the coordinate ${\bf r}$ given by
\begin{eqnarray}
 \Sigma_1({\bf r})=e^{-{ i \over 2} \chi({\bf r})} 
 \left(
 \begin{array}{c}
 e^{-i { 1 \over 2} \xi ({\bf r})} \sin {{\zeta ({\bf r}) } \over 2}
 \\
  e^{i { 1 \over 2} \xi {\bf r})} \cos {{\zeta ({\bf r}) } \over 2} 
 \end{array}
 \right)
 \label{spin-d1}
\end{eqnarray}
where $\zeta ({\bf r})$ and $\xi ({\bf r})$ are the polar and azimuthal angles of the spin-direction, respectively.
This coordinate dependence is necessary to describe the spin-twisting itinerant motion.
The expectation value of spin ${\bf s}({\bf r})=(s_x({\bf r}), s_y({\bf r}), s_z({\bf r}))$ is given by
\begin{eqnarray}
s_x ({\bf r})= { \hbar \over 2} \cos \xi ({\bf r}) \sin \zeta ({\bf r}), \ s_y ({\bf r})= { \hbar \over 2} \sin \xi ({\bf r}) \sin \zeta ({\bf r}), \ s_z= { \hbar \over 2}  \cos \zeta ({\bf r})
\end{eqnarray}

In the following argument, the single-valued requirement of the wave function as a function of the coordinate is a crucial condition.
This is the postulate adopted by Schr\"{o}dinger \cite{Schrodinger}, and we must impose this condition on the wave packet.

We consider the case where the electron performs spin-twisting itinerant motion in which $\xi ({\bf r}) \rightarrow \xi ({\bf r})+2\pi$ occurs after a circular transport along a loop in the coordinate space.
The angular variable $\chi({\bf r})$ in Eq.~(\ref{spin-d1}) is introduced to
make the spin-function single valued; without it the spin function becomes multi-valued for the shift $\xi ({\bf r}) \rightarrow \xi ({\bf r})+2\pi$
due to the phase factors $e^{\pm i { 1 \over 2} \xi ({\bf r})}$ in Eq.~(\ref{spin-d1}), resulting the wave packet multi-valued. The condition on $\chi$ 
to impose the single-valued requirement will be given, later.

The distribution function $a({\bf q},t)$ satisfies the normalization 
\begin{eqnarray}
 \int d^3{q} \, |a({\bf q},t)|^2&=&1
 \end{eqnarray}
 and the localization condition in ${\bf k}$ space,
 \begin{eqnarray}
 \int d^3{q} \, {\bf q} |a({\bf q},t)|^2&=&{\bf q}_c
\end{eqnarray}
The distribution of $|a({\bf q},t)|^2$ is assumed to be narrow compared with the Brillouin zone size so that ${\bf q}_c$ can be regarded as the central wave vector of the wave packet.

The wave packet is also localized in ${\bf r}$ space around the central position ${\bf r}_c$,
\begin{eqnarray}
{\bf r}_c&=& \langle ({\bf q}_c, {\bf r}_c) |{\bf r} |({\bf q}_c, {\bf r}_c) \rangle.
\end{eqnarray}
The localization in the ${\bf q}$ and ${\bf r}$ spaces is assumed to satisfy the Heisenberg uncertainty principle \cite{DiracWP}.

 We include the following Rashba interaction term in the Hamiltonian
\begin{eqnarray}
H_{so}= {\bm \lambda}({\bf r})\cdot {{\hbar{\bm \sigma}} \over 2} \times \left(\hat{\bf p}-{q}{\bf A}^{\rm em}({\bf r})\right), 
\end{eqnarray}
where ${\bm \lambda}({\bf r})$ is the spin-orbit coupling vector (its direction is the internal electric field direction), $\hat{\bf p}=-i\hbar \nabla$ is the momentum operator, and $q=-e$ is electron charge \cite{Rashba}. 

Let us construct the Lagrangian $L'({\bf r}_c, \dot{\bf r}_c,{\bf q}_c,\dot{\bf q}_c)$ using the time-dependent variational principle \cite{Koonin1976}, 
\begin{eqnarray}
L'=\langle ({\bf q}_c, {\bf r}_c) | i \hbar {\partial \over {\partial t}}-H| ({\bf q}_c, {\bf r}_c) \rangle.
\end{eqnarray}
where $H$ is composed of the Hamiltonian for the band electron that gives the band dispersion ${\cal E}\left({\bf q}_c+{ e \over {\hbar}} {\bf A}_1^{\rm eff}({\bf r}_c)\right)$ and the Rashba interaction $H_{so}$.

For convenience sake, we introduce another Lagrangian $L$ that is related to $L'$ as 
\begin{eqnarray}
L=L'-\hbar{d \over {dt}}\left[ \gamma({\bf q}_c,t) -{\bf r}_c \cdot {\bf q}_c \right],
\end{eqnarray}
where $\gamma$ is the phase of $a({\bf q},t)=|a({\bf q},t)|e^{-i \gamma({\bf q},t)}$.

By following  procedures for calculating expectation values for operators by the wave packet \cite{Niu}, 
$L$ is obtained as 
\begin{eqnarray}
L&=&-{\cal E}\left({\bf q}_c+{ e \over {\hbar}} {\bf A}_1^{\rm eff}({\bf r}_c)\right) 
+\hbar {\bf q}_c \cdot \dot{{\bf r}_c}+i\hbar \left\langle u_{\bf q} \left| {{d u_{\bf q} } \over {dt } } \right. \right\rangle
\nonumber
\\
&+&\hbar {\bm \lambda}({\bf r}_c)\cdot \left[{\bf s({\bf r}_c)} \times \left({\bf q}_c+{ e \over {\hbar}} {\bf A}_1^{\rm eff}({\bf r}_c) \right)\right],
\end{eqnarray}
 where ${\bf s}({\bf r}_c)$ is the expectation value of spin for the wave packet centered at ${\bf r}_c$ given by
 \begin{eqnarray}
{\bf s}({\bf r}_c)={\hbar \over 2} \langle ({\bf q}_c, {\bf r}_c) | {\bm \sigma} |({\bf q}_c, {\bf r}_c)\rangle.
\end{eqnarray}
and
\begin{eqnarray}
{\bf A}_1^{\rm eff} ={\bf A}^{\rm em}+ {\bf A}_1^{\rm fic}
\label{Aeff}
\end{eqnarray}
where ${\bf A}_1^{\rm fic}$ is the Berry connection arising from $\Sigma_1$ given by
\begin{eqnarray}
{\bf A}^{\rm fic}_1({\bf r})=-i {\hbar \over e}\Sigma_1^{\dagger} \nabla \Sigma_1= -{\hbar \over {2e}} \nabla \chi ({\bf r}) +{ \hbar \over {2e}} \nabla \xi ({\bf r}) \cos \zeta ({\bf r})
\end{eqnarray}

We introduce the following wave vector ${\bf k}_c$, 
\begin{eqnarray}
{\bf k}_c = {\bf q}_c+{ e \over {\hbar}} {\bf A}^{\rm eff}_1({\bf r}_c)
\label{gaugek}
\end{eqnarray}
and change the dynamical variables from ${\bf q}_c, \dot{\bf q}_c$ to ${\bf k}_c, \dot{\bf k}_c$ \cite{Niu}.
 
Then, the Lagrangian with dynamical variables ${\bf r}_c, \dot{\bf r}_c,{\bf k}_c,\dot{\bf k}_c$ is given by
\begin{eqnarray}
&&L({\bf r}_c, \dot{\bf r}_c,{\bf k}_c,\dot{\bf k}_c)=-{\cal E}({\bf k}_c) +\hbar {\bm \lambda}({\bf r}_c)\cdot \left[{\bf s}({\bf r}_c) \times {\bf k}_c \right]
\nonumber
\\
&+&\hbar \left[ {\bf k}_c -{ e \over {\hbar}} {\bf A}^{\rm eff}({\bf r}_c) \right] \cdot \dot{{\bf r}_c}
+i\hbar \dot{{\bf k}}_c \cdot \left \langle u_{\bf q} | {{\partial u_{\bf q} } \over {\partial {\bf q} } } \right\rangle_{ {\bf q}={\bf k}_c}
\label{Lag}
\end{eqnarray}

Using the above Lagrangian $L$, the following equations of motion are obtained,
\begin{eqnarray}
\dot{\bf r}_c&=&{ 1 \over \hbar} {{\partial {\cal E}} \over {\partial {\bf k}_c}}+ {\bm \lambda}({\bf r}_c) \times {\bf s}({\bf r}_c)-\dot{\bf k}_c\times {\bm \Omega},
\label{eqm1}
\\
\dot{\bf k}_c&=&   { {  \partial  } \over {\partial {\bf r}_c}}\left[{\bm \lambda}({\bf r}_c) \times{\bf s}({\bf r}_c) \cdot {\bf k}_c  \right] -{e \over {\hbar }}\dot{\bf r}_c\times {\bf B}^{\rm eff},
\label{eqm2}
\end{eqnarray}
where ${\bm \Omega}$ is the Berry curvature in ${\bf k}$ space defined by
\begin{eqnarray}
{\bm \Omega}=i\hbar\nabla_{\bf q} \times  \left\langle u_{\bf q} | \nabla_{\bf q}| u_{\bf q} \right\rangle
\end{eqnarray}
and ${\bf B}^{\rm eff}$ is the effective magnetic field,
\begin{eqnarray}
{\bf B}^{\rm eff}=\nabla \times {\bf A}_1^{\rm eff}
\end{eqnarray}

In the following, we consider the case where ${\bm \Omega}=0$. 
Then, Eq.~(\ref{eqm1}) becomes
\begin{eqnarray}
\dot{\bf r}_c={ 1 \over \hbar} {{\partial {\cal E}({\bf k}_c)} \over {\partial {\bf k}_c}}+{\bm \lambda}({\bf r}_c)\times {\bf s}({\bf r}_c).
\label{eqm3}
\end{eqnarray}

Using Eq.~(\ref{eqm3}), Eq.~(\ref{eqm2}) becomes,
\begin{eqnarray}
\dot{\bf k}_c &=&  { {  \partial  } \over {\partial {\bf r}_c}}\left[\left(\dot{\bf r}_c-{ 1 \over \hbar} {{\partial {\cal E}({\bf k}_c)} \over {\partial {\bf k}_c}}\right) \cdot {\bf k}_c  \right] -{e \over {\hbar}}\dot{\bf r}_c\times {\bf B}^{\rm eff}
\nonumber
\\
&=&-{e \over {\hbar}}\dot{\bf r}_c\times {\bf B}^{\rm eff}
\label{eqm4}
\end{eqnarray}

 Eqs.~(\ref{eqm3}) and (\ref{eqm4}) indicate that the wave packet exhibits cyclotron motion for the electron in the band with energy
\begin{eqnarray}
{\cal E}({\bf k})+\hbar{\bm \lambda}({\bf r})\times {\bf s}({\bf r})\cdot {\bf k}
\label{NewB}
\end{eqnarray}

By following the Onsager's argument, let us quantize the cyclotron orbit \cite{Onsager1952}. From Eq.~(\ref{Lag}), the Bohr-Sommerfeld relation becomes
\begin{eqnarray}
\oint_C (\hbar {\bf k}_c -e{\bf A}_1^{\rm eff}) \cdot d{\bf r}_c =2\pi \hbar \left(n+{ 1 \over 2} \right)
\label{Onsager1}
\end{eqnarray}
where $n$ is an integer and $C$ is the closed loop that corresponds to the section of Fermi surface enclosed by the cyclotron motion.

Using Eq.~(\ref{eqm4}), we have 
\begin{eqnarray}
\oint_C \hbar {\bf k}_c \cdot d{\bf r}_c&=&-e \oint_C d{\bf r}_c \cdot {\bf r}_c \times {\bf B}^{\rm eff}=e \oint_C  {\bf B}^{\rm eff} \cdot  {\bf r}_c \times d{\bf r}_c
\nonumber
\\
&=&2e \oint_C  {\bf A}_1^{\rm eff} \cdot d{\bf r}_c
\end{eqnarray}

Thus, Eq.~(\ref{Onsager1}) becomes
\begin{eqnarray}
e \oint_C  {\bf A}^{\rm em} \cdot  d{\bf r}_c+e \oint_C  {\bf A}_1^{\rm fic} \cdot  d{\bf r}_c=2\pi \hbar \left(n+{ 1 \over 2} \right)
\label{Onsager2}
\end{eqnarray}
Note that in the usual quantization condition, ${\bf A}_1^{\rm fic}$ is absent, and the above becomes the quantized condition for the cyclotron motion.

Now we consider the case where ${\bf A}_1^{\rm fic}$ is present.
The above quantization condition is satisfied even if the magnetic field is absent. 
If the magnetic field is zero, the first term is zero, and we have
\begin{eqnarray}
- \oint_C {1 \over 2} \nabla \chi ({\bf r})\cdot  d{\bf r}_c + \oint_C { 1 \over 2} \nabla \xi ({\bf r}) \cos \zeta ({\bf r}) \cdot  d{\bf r}_c=2\pi \left(n+{ 1 \over 2} \right)
\end{eqnarray}
which has two solutions, one is $\zeta=\pi/2$, $w_C[\chi]=-1, n=0$; and the other is $\zeta=\pi/2$, $w_C[\chi]=1, n=-1$, where
\begin{eqnarray}
w_C[\chi]= {1 \over {2\pi}} \oint_C\nabla \chi ({\bf r})\cdot  d{\bf r}
\end{eqnarray}
is the winding number of $\chi$ along loop $C$. We will argue later that the condition $\zeta=\pi/2$ may be achieved by
the kinetic energy gain if electron pairs are formed.

The above solutions correspond to the case where ${\bf A}_1^{\rm fic}$ is given by
\begin{eqnarray}
{\bf A}^{\rm fic}=-{\hbar \over {2e}} \nabla \chi
\label{Afic}
\end{eqnarray}

The condition $\zeta=\pi/2$ leads to the following requirements
\begin{eqnarray}
w_C[\chi]+w_C[\xi]= \mbox{even number}
\end{eqnarray}
from the single-valuedness condition for the spin function $\Sigma_1$ as a function of the coordinate ${\bf r}$; if the above is satisfied, the phase factors $e^{-{ i \over 2} \chi({\bf r})}  e^{\pm i { 1 \over 2} \xi ({\bf r})}$ in Eq.~(\ref{spin-d1}) become single-valued.

The condition $w_C[\chi]=\pm1$ requires that $w_C[\xi]$ must be odd, thus, $w_C[\xi]$ is not zero. The nonzero value of $w_C[\xi]$ means that electrons perform spin-twisting itinerant motion. This indicates that the quantized cyclotron motion may occur without an external magnetic field when the itinerant motion is accompanied by the spin-twisting.

\section{The pairing energy gap}
\label{section6}

The results in the previous section indicate that due to the presence of the Rashba interaction, the band energy becomes the one  in Eq.~(\ref{NewB}), and
Bloch electrons may perform the spin-twisting itinerant motion. 

In this section, we consider a modified BCS model where the pairing between single particle states $({\bf k}, {\bf s}_0({\bf r}))$ and $(-{\bf k}, -{\bf s}_0({\bf r}))$ occurs, 
instead between $({\bf k}, \uparrow)$ and $(-{\bf k}, \downarrow)$; the use of ${\bf s}_0({\bf r})$ enables to take into account the possibility for the occurrence of the spin-twisting itinerant motion. 

Since we use the results of the BCS theory below, let us briefly review it first\cite{BCS1957}. The Hamiltonian for the BCS model is given by $H_{\rm kin}+H_{\rm int}$, where $H_{\rm kin}$ is the kinetic energy given by
\begin{eqnarray}
H_{\rm kin}= \sum_{{\bf k} \sigma} \xi_0({\bf k})c^{\dagger}_{{\bf k} \sigma}c_{{\bf k} \sigma}
\end{eqnarray}
$\xi({\bf k})$ is the energy measured from the Fermi energy $\mu$ given by
\begin{eqnarray}
\xi_0({\bf k})={\cal E}({\bf k})-\mu
\end{eqnarray}
and $H_{\rm int}$ is the interaction energy given by
\begin{eqnarray}
H_{\rm int}={1 \over 2} \sum_{{\bf k} {\bm \ell}} V_{{\bf k} {\bm \ell}}c^{\dagger}_{{\bf k} \uparrow}c^{\dagger}_{-{\bf k} \downarrow}c_{-{\bm \ell} \downarrow}c_{{\bm \ell} \uparrow}.
\label{Hint}
\end{eqnarray}
The electron pairing occurs between electrons near the Fermi surface due to an attractive $V_{{\bf k} {\bm \ell}}$ that exists in that region.
In the BCS model, $V_{{\bf k} {\bm \ell}}$ is nonzero ($V_{{\bf k} {\bm \ell}}=-g$) only when $|{\xi}_0({\bf k})|, |{\xi}_0({\bm \ell})| <  \hbar \omega_D$ ($\omega_D$
 is the Debye frequency) is satisfied. Then, $\Delta_{\bf k}$ becomes independent of ${\bf k}$, will be expred as $\Delta$.

The superconducting state is given by the following state vector,
\begin{eqnarray}
|{\rm BCS} \rangle=\prod_{\bf k}(u_{\bf k}+v_{\bf k}c^{\dagger}_{{\bf k} \uparrow}c^{\dagger}_{-{\bf k} \downarrow})
|{\rm vac} \rangle.
\label{BCS}
\end{eqnarray}
This state exploits the attractive interaction between electron pairs $({\bf k} \uparrow)$ and $(-{\bf k} \downarrow)$ and the following energy gap equation is 
obtained,
\begin{eqnarray}
\Delta=g\sum_{ |{\xi}_0({\bm \ell})| <  \hbar \omega_D } u_{\bm \ell}v_{\bm \ell}
\end{eqnarray}
 and
 $u_{\bf k}$ and $v_{\bf k}$ are parameters given using $\Delta$ and $\xi({\bf k})_0$ as
\begin{eqnarray}
u_{\bf k}={1 \over \sqrt{2}} \left(1 + {{\xi_0({\bf k})} \over \sqrt{\xi_0^2({\bf k})+\Delta^2}} \right)^{1/2}
\end{eqnarray}
and
\begin{eqnarray}
v_{\bf k}={1 \over \sqrt{2}} \left(1 -{{\xi_0({\bf k})} \over \sqrt{\xi_0^2({\bf k})+\Delta^2}} \right)^{1/2},
\end{eqnarray}
respectively.

The total energy by the formation of the energy gap is given by
\begin{eqnarray}
E_{\rm s}^{\rm BCS}&=&E_{\rm n}^{\rm BCS}-{1 \over 2} N(\mu)\Delta^2
\end{eqnarray}
 where $E_{\rm n}^{\rm BCS}$ is the normal state energy, and $N(\mu)$ is the density of states at the Fermi energy $\mu$ \cite{BCS1957}.

Now, consider the pairing of $({\bf k}, {\bf s}_0({\bf r}))$ and $(-{\bf k}, -{\bf s}_0({\bf r}))$, and also $({\bf k}, -{\bf s}_0({\bf r}))$ and $(-{\bf k}, {\bf s}_0({\bf r}))$. We divide the system into coarse-grained cells of volume $1$ to take into account the coordinate dependence of the band energy in Eq.~(\ref{NewB}), assuming that its coordinate dependence is very slow.
 Then, the ground state in the cell with the central position ${\bf r}$ is given by
\begin{eqnarray}
|{\rm BCS} ({\bf r}) \rangle=\prod_{\bf k}(u_{\bf k}({\bf r})+v_{\bf k}({\bf r})c^{\dagger}_{{\bf k} \uparrow}c^{\dagger}_{-{\bf k} \downarrow})
|{\rm vac} \rangle.
\label{BCS}
\end{eqnarray}
where  $u_{\pm}({\bf k}, {\bf r})$ and $v_{\pm}({\bf k}, {\bf r})$ are given by
\begin{eqnarray}
u_{\pm}({\bf k}_c, {\bf r})&=&{1 \over \sqrt{2}} \left(1 + {{{\xi}_{\pm}({\bf k}, {\bf r})} \over \sqrt{{\xi}^2_{\pm}({\bf k}, {\bf r})+ \Delta^2( {\bf r})}} \right)^{1/2},
\nonumber
\\
&&
\\
v_{\pm}({\bf k}, {\bf r})&=&{1 \over \sqrt{2}} \left(1 - {{{\xi}_{\pm}({\bf k}, {\bf r})} \over \sqrt{{\xi}^2_{\pm}({\bf k}, {\bf r})+ \Delta^2 ({\bf r})}} \right)^{1/2},
\nonumber
\\
\end{eqnarray}
with
\begin{eqnarray}
\xi_{\pm}({\bf k}, {\bf r})={\cal E}_{\pm}({\bf k})-\mu, \quad {\cal E}_{\pm} ({\bf k}, {\bf r})={\cal E}({\bf k}) \pm\hbar {\bm \lambda} ({\bf r}) \times {\bf k} \cdot {\bf s}_0({\bf r})
\label{Epm}
\end{eqnarray}

Then, the gap function $\Delta( {\bf r})$ is the solution of the gap equation given by
\begin{eqnarray}
\Delta({\bf r})\!&=&\!{{g} \over 2}\sum_{|{\xi}_{0}({\bm \ell})| < \hbar \omega_D }  \left\{ u_{+}({\bm \ell }, {\bf r})v_{+}({\bm \ell}, {\bf r})
\!+\!u_{-}({\bm \ell}, {\bf r})v_{-}({\bm \ell}_c, {\bf r}) \right\}
\nonumber
\\
\!&=&{{g \Delta({\bf r})} \over 4}\!\sum_{|{\xi}_{0}({\bm \ell})| < \hbar \omega_D }  \left\{ {{1} \over \sqrt{{\xi}^2_{+}({\bf k}, {\bf r})+ \Delta^2 ({\bf r}_c)}} 
\!+\! {{1} \over \sqrt{{\xi}^2_{-}({\bf k}, {\bf r})+ \Delta^2 ({\bf r})}}   \right\}
\nonumber
\\
\!&\approx&{{g \Delta({\bf r})} \over 4}\!\sum_{|{\xi}_{0}({\bm \ell})| < \hbar \omega_D }  \left\{ {{2} \over \sqrt{{\xi}^2_{0}({\bf k}, {\bf r})+ \Delta^2 ({\bf r})}} 
\!-\! {{\lambda^2} \over {[{\xi}^2_{0}({\bf k}, {\bf r})+ \Delta^2 ({\bf r})]^{3/2}}}   \right\}
\nonumber
\\
\!&\approx&{{g \Delta({\bf r}) N(\mu;{\bf r})} \over 4} \int_{-\hbar \omega_D}^{\hbar \omega_D} d \epsilon  \left\{ {{2} \over \sqrt{\epsilon^2+ \Delta^2 ({\bf r})}} 
\!-\! {{\lambda^2} \over {[{\epsilon}^2+ \Delta^2 ({\bf r})]^{3/2}}}   \right\}
\nonumber
\\
\!&\approx&{{g \Delta({\bf r}) N(\mu;{\bf r})}} \left\{ \log {{2 \hbar \omega_D} \over {\Delta({\bf r})}} 
\!-\! {{\lambda^2} \over {\Delta^2({\bf r})}}   \right\}
\label{gap}
\end{eqnarray}
where $N(\mu;{\bf r})$ is the density of states at the Fermi energy in the coarse grained cell with the central position ${\bf r}$.

From the above relation, we can obtained the following,
\begin{eqnarray}
\Delta({\bf r}) \approx  2 \hbar \omega_D \exp \left( - { 1 \over {g N(\mu;{\bf r})}} -{ {\lambda^2}  \over {\Delta_0^2}} \right ); \quad \Delta_0({\bf r}) = 2 \hbar 
\omega_D \exp \left( - { 1 \over {g N(\mu;{\bf r})}} \right)
 \label{gap2}
 \end{eqnarray}
 where $\Delta_0$ is the gap value without the spin-orbit interaction; here, it is assumed that $\hbar \omega_D \gg \Delta$ holds. 
 The gap $\Delta$ is reduced by the spin-orbit interaction. However, if the spin-orbit interaction parameter $\lambda$ is significantly smaller that $\Delta_0$, the gap is almost the same as the original one.
 In the following we assume such a case. 
 
\section{Kinetic energy gain by the spin-twisting itinerant motion}
\label{section8}
 
In this section, we consider the appearance of ${\bf A}^{\rm fic}=-{\hbar \over {2e}} \nabla \chi$ in Eq.~(\ref{Afic}) from the view point of the kinetic energy gain.

We consider the case where the pair $({\bf k}, {\bf s}_0({\bf r}))$ and $(-{\bf k}, -{\bf s}_0({\bf r}))$, and another pair $({\bf k}, -{\bf s}_0({\bf r}))$ and $(-{\bf k}, {\bf s}_0({\bf r}))$, are both occupied. 
We assume that ${\bf s}_0({\bf r})$ for the first pair arises from the spin function $\Sigma_1$ in Eq.~(\ref{spin-d1}), and $- {\bf s}_0({\bf r})$ for the second pair arises from the spin function $\Sigma_2$ given by
\begin{eqnarray}
 \Sigma_2({\bf r})=e^{-{ i \over 2} \chi({\bf r})} 
 \left(
 \begin{array}{c}
 ie^{-i { 1 \over 2} \xi ({\bf r})} \cos {{\zeta({\bf r})} \over 2}
 \\
 -i e^{i { 1 \over 2} \xi {\bf r})} \sin {{\zeta({\bf r})} \over 2}
 \end{array}
 \right)
 \end{eqnarray}
 Note that  $\Sigma_1$ and $\Sigma_2$ are orthogonal.
 
The fictitious vector potential ${\bf A}_2^{\rm fic}({\bf r})$ from $\Sigma_2$ is calculated as
\begin{eqnarray}
{\bf A}_2^{\rm fic}({\bf r})=-i {\hbar \over e}\Sigma_2^{\dagger} \nabla \Sigma_2= -{\hbar \over {2e}} \nabla \chi ({\bf r}) -{ \hbar \over {2e}} \nabla \xi ({\bf r}) \cos \zeta ({\bf r})
\end{eqnarray}
 and the effective vector potential is given by
 \begin{eqnarray}
{\bf A}_2^{\rm eff} ={\bf A}^{\rm em}+ {\bf A}_2^{\rm fic}
\end{eqnarray}

 The single-particle energy for the pair $({\bf k}, {\bf s}_0({\bf r}))$ and $(-{\bf k}, -{\bf s}_0({\bf r}))$ is $ {\cal E}_{+}({\bf k}, {\bf r})$, and that for the pair $({\bf k}, -{\bf s}_0({\bf r}))$ and $(-{\bf k}, {\bf s}_0({\bf r}))$  is $ {\cal E}_{-}({\bf k}, {\bf r})$ in Eq.~(\ref{Epm}).

Then, the kinetic energy for the cell at ${\bf r}$ is given by
\begin{eqnarray}
 2\sum_{ {\cal E}_-({\bf k}, {\bf r}) < \mu}  {\cal E}_-({\bf k}, {\bf r})+2\sum_{ {\cal E}_+({\bf k}, {\bf r}) < \mu}   {\cal E}_+({\bf k}, {\bf r})
 \label{Ekin}
\end{eqnarray}

For simplicity, we approximate the above energy using the Fermi distribution functions $f(\epsilon)=(1+ e^{\epsilon/k_BT})^{-1}$ ($k_B$ is Boltzmann's constant) and density of states $N(\epsilon;{\bf r})$ as
\begin{eqnarray}
&&\int  {{N(\epsilon;{\bf r})} \over 2} \Big\{ [\epsilon+ \hbar {\bm \lambda}({\bf r}) \times {\bf k}_c\cdot {\bf s}_0({\bf r})]
f(\epsilon+ \hbar {\bm \lambda}({\bf r}) \times {\bf k}_c\cdot {\bf s}_0({\bf r}))
\nonumber
\\
&+&
[\epsilon- \hbar {\bm \lambda}({\bf r}) \times {\bf k}_c\cdot {\bf s}_0({\bf r})]
f(\epsilon-\hbar {\bm \lambda}({\bf r}) \times {\bf k}_c\cdot {\bf s}_0({\bf r}))
 \Big\}d \epsilon
\nonumber
 \\
  &\approx&
  \int  {{N(\epsilon;{\bf r})} \over 2} \Big\{ \epsilon \left[f(\epsilon+ \hbar {\bm \lambda}({\bf r}) \times {\bf k}\cdot {\bf s}_0({\bf r}))+f(\epsilon- \hbar {\bm \lambda}({\bf r}) \times {\bf k} \cdot {\bf s}_0({\bf r})) \right]
  \nonumber
\\
  &+& \hbar {\bm \lambda}({\bf r}) \times {\bf k}\cdot {\bf s}_0({\bf r}) \left[f(\epsilon+ \hbar {\bm \lambda}({\bf r}) \times {\bf k}\cdot {\bf s}_0({\bf r}))-f(\epsilon- \hbar {\bm \lambda}({\bf r}) \times {\bf k}\cdot {\bf s}_0({\bf r})) \right] \Big\} d \epsilon
\nonumber
\\
 &\approx &
  \int  {{N(\epsilon;{\bf r})} \over 2} \left\{ 2 \epsilon f(\epsilon) 
  + 2\left|\hbar {\bm \lambda}({\bf r}) \times {\bf k}\cdot {\bf s}_0({\bf r}) \right|^2 {{\partial f(\epsilon)} \over {\partial \epsilon}}  \right\}d \epsilon
\label{energy0}
\end{eqnarray}

At temperature $T=0$,  ${{\partial f(\epsilon)} \over {\partial \epsilon}} =- \delta( \epsilon)$; thus, we have
\begin{eqnarray}
  \int d \epsilon N(\epsilon;{\bf r})  \epsilon f(\epsilon) d \epsilon- N(\mu;{\bf r})  \left|\hbar {\bm \lambda}({\bf r}) \times {\bf k}_c\cdot {\bf s}_0({\bf r}) \right|^2 
  \label{energy0a}
\end{eqnarray}

The first term in Eq.~(\ref{energy0a}) may be approximated as
\begin{eqnarray}
 {1 \over 2} \sum_{{\cal E} ({\bf q}) < \mu, i=1,2}{ {\hbar^2} \over {2m}}\left[ {\bf q}+{e \over \hbar } {\bf A}_i^{\rm eff} \right]^2  \approx
  \sum_{{\cal E} ({\bf q}) < \mu}
  { {\hbar^2} \over {2m}}{\bf q}^2+{{e^2 \rho({\bf r})} \over {4m}}( |{\bf A}_1^{\rm eff}|^2 +|{\bf A}_2^{\rm eff}|^2 )
    \label{energy0ab}
\end{eqnarray}
assuming that the term linear in ${\bf q}$ cancels out due to the time-reversal and/or inversion symmetry. Here $\rho$ is the number density of electrons (later, we consider it as the number density of electrons participating in the collective mode $\nabla \chi$).

The second term in Eq.~(\ref{energy0a}) may be approximated as
\begin{eqnarray}
&&  - { 1 \over 2}\sum_{{\cal E}({\bf q}) =\mu, j=1,2}\left|\hbar {\bm \lambda}({\bf r}) \times \left[ {\bf q}+{e \over \hbar } {\bf A}_j^{\rm eff} \right]\cdot {\bf s}_0({\bf r}) \right|^2 
    \nonumber
    \\
     &\approx&  - \hbar^2\sum_{{\cal E}({\bf q}) =\mu}\left| {\bm \lambda}({\bf r}) \times {\bf q}\cdot {\bf s}_0({\bf r}) \right|^2 
      -{ 1 \over 2}\sum_{j=1,2} e^2 N(\mu;{\bf r})\left| {\bm \lambda}({\bf r}) \times {\bf s}_0({\bf r}) \cdot {\bf A}_j^{\rm eff}\right|^2   
  \label{energy0b}
\end{eqnarray}
assuming that the term linear in ${\bf q}$ cancels out.

From the condition for minimizing the kinetic energy, ${\bf s}_0$ is chosen to satisfy
\begin{eqnarray}
{\bm \lambda ({\bf r})} \times {\bf s}_0({\bf r})  \parallel {\bf A}_1^{\rm eff}({\bf r}) \mbox{ and }  {\bf A}_2^{\rm eff}({\bf r}) 
\label{conds0}
\end{eqnarray}

We assume ${\bm \lambda ({\bf r})}$ in the coarse-grained cell at ${\bf r}$ to be uniform in the $z$-direction; then, the optimal ${\bf s}_0({\bf r})$ that satisfies the above condition lies in the $xy$ plane. Thus, 
$\zeta$ in ${\bf A}_1^{\rm eff}({\bf r})$ and ${\bf A}_2^{\rm eff}({\bf r})$ is taken to be $\zeta=\pi/2$, yielding
${\bf A}_1^{\rm fic}({\bf r})={\bf A}_2^{\rm fic}({\bf r})=-{\hbar \over {2e}} \nabla \chi$. 
As a consequence, we have the common effective potential for ${\bf A}_1^{\rm eff}({\bf r})$ and ${\bf A}_2^{\rm eff}({\bf r})$ given by
\begin{eqnarray}
{\bf A}^{\rm eff}={\bf A}^{\rm em}-{\hbar \over {2e}} \nabla \chi={\bf A}^{\rm em}+{\bf A}^{\rm fic}
\end{eqnarray}

The kinetic energy increase given by the appearance of ${\bf A}^{\rm eff}$ in Eq.~(\ref{energy0ab}) is calculated as 
 \begin{eqnarray}
 \int d^3 r  {{e^2 \rho({\bf r})} \over {2m}} |{\bf A}^{\rm eff}|^2 
 \end{eqnarray}
 This indicates that the optimum ${\bf A}^{\rm fic}$ is the one that gives ${\bf A}^{\rm eff}=0$ if this choice is possible. 
 If we adopt ${\bf A}^{\rm em}=0$ when a magnetic field is zero, the condition yields ${\bf A}^{\rm fic}=0$, i.e., the absence of the spin-twisting itinerant motion.
  When ${\bf A}^{\rm em} \neq 0$, however, the optimal ${\bf A}^{\rm fic}$ will be the one for the presence of the spin-twisting itinerant motion.

 From the kinetic energy, we can calculate the current density as
\begin{eqnarray}
{\bf j}_{\rm tot}({\bf r})=- e^2  \left[{{\rho ({\bf r})} \over {m}} - N(\mu; {\bf r}) |{\bm \lambda} ({\bf r})\times {\bf s}_0 ({\bf r})|^2 \right]
{\bf A}^{\rm eff}({\bf r})
\label{current3}
\end{eqnarray}
This is the London equation, and the system should exhibit the Meissner effect. Thus, ${\bf A}^{\rm eff}=0$ is realized in the bulk. 
If the system is a ring-shaped, it will lead to the quantization of magnetic flux in the units ${ h \over {2e}}$. 
The equation (\ref{energy0b}) indicates the occurrence of the energy reduction in the order of $\lambda^2$ if the surface energy is negligible compared to the bulk energy. 
In other words, when a magnetic field is applied the spin-twisting itinerant motion occurs, and gives rise to ${\bf A}^{\rm fic}$ that causes the Meissner effect and the flux quantization  in ${ h \over {2e}}$.

 \section{Berry connection for many-body wave functions and ${\bf A}^{\rm fic}$ }
\label{section2}

We consider ${\bf A}^{\rm fic}$ from the view point of the {\em Berry connection for the many-body wave functions} (dented as ${\bf A}^{\rm MB}$) introduced in our previous work \cite{koizumi2019} in this section.
 
Let us denote the wave function of a system with $N$ electrons as
\begin{eqnarray}
\Psi ({\bf x}_1, \cdots, {\bf x}_{N},t)
\label{wavef}
\end{eqnarray}
where ${\bf x}_j=({\bf r}_j,s_j) $ denotes the coordinate ${\bf r}_j$  and spin $s_j$ of the $j$th electron.

The Berry connection\cite{Berry} associated with this wave function is called the ``{\em Berry connection for many-body wave function'' \cite{koizumi2019}. 
In order to calculate this Berry connection}, we first prepare the parameterized wave function $|n_{\Psi}({\bf r}) \rangle$ with the parameter ${\bf r}$, 
 \begin{eqnarray}
\langle s, {\bf x}_{2}, \cdots, {\bf x}_{N} |n_{\Psi}({\bf r},t) \rangle = { {\Psi({\bf r}s, {\bf x}_{2}, \cdots, {\bf x}_{N},t)} \over {|C({\bf r} ,t)|^{{1 \over 2}}}}
\end{eqnarray}
where $|C({\bf r} ,t)|$ is the normalization constant given by 
\begin{eqnarray}
|C({\bf r} ,t)|=\int ds d{\bf x}_{2} \cdots d{\bf r}_{N}\Psi({\bf r} s, {\bf x}_{2}, \cdots)\Psi^{\ast}({\bf x} s, {\bf x}_{2}, \cdots)
\end{eqnarray}

Using $|n_{\Psi}\rangle$, the {\em Berry connection for many-body wave function} is given by
 \begin{eqnarray}
{\bf A}^{\rm MB}({\bf r},t)=-i \langle n_{\Psi}({\bf r},t) |\nabla_{\bf r}  |n_{\Psi}({\bf r},t) \rangle
\end{eqnarray}
Here, ${\bf r}$ is regarded as the parameter \cite{Berry}. 

When the origin of ${\bf A}^{\rm MB}$ is not the ordinary magnetic field one, i.e.,  
\begin{eqnarray}
\nabla \times {\bf A}^{\rm MB}=0
\label{BMB}
\end{eqnarray}
 it can be written in the pure gauge form,
\begin{eqnarray}
 {\bf A}^{\rm MB}=\nabla \theta
\end{eqnarray}
where $\theta$ is a function which may be multi-valued.

Let us consider the case where $\Psi$ is given as a Slater determinant
of spin-orbitals $\phi_{1,1}({\bf r})\Sigma_1({\bf r}), \phi_{1,2}({\bf r})\Sigma_2({\bf r})$, $\dots$, $\phi_{{N \over 2},1}({\bf r})\Sigma_1({\bf r})$, and $\phi_{{N \over 2},2}({\bf r})\Sigma_2({\bf r}) $,
where $\phi_{j,1}({\bf r})$ and $\phi_{j,2}({\bf r})$ are time-reversal partners and $N$ is assumed to be even. 

Then, ${\bf A}^{\rm MB}$ is calculated as
\begin{eqnarray}
 {\bf A}^{\rm MB}&=& \  \rm{Im}
{ { \sum_{j=1}^{N \over 2} \left[ \phi^{\ast}_{j,1}({\bf r})\Sigma^{\dagger}_1({\bf r}) \nabla \phi_{j,1}({\bf r})\Sigma_1({\bf r})+\phi^{\ast}_{j,2}({\bf r})\Sigma^{\dagger}_2({\bf r})\nabla 
 \phi_{j,2}({\bf r})\Sigma_2({\bf r}) \right]} \over
  { \sum_{j=1}^{N \over 2} \left[ \phi^{\ast}_{j,1}({\bf r}) \phi_{j,1}({\bf r})+\phi^{\ast}_{j,2}({\bf r})
 \phi_{j,2}({\bf r}) \right] } }
 \nonumber
 \\
 &=&
 \  { e \over \hbar}
{ {{\bf A}^{\rm fic}_1\sum_{j=1}^{N \over 2}\phi^{\ast}_{j,1}({\bf r})
 \phi_{j,1}({\bf r}) + {\bf A}^{\rm fic}_2\sum_{j=1}^{N \over 2}\phi^{\ast}_{j,2}({\bf r})
 \phi_{j,2}({\bf r})} \over
  { \sum_{j=1}^{N \over 2} \left[ \phi^{\ast}_{j,1}({\bf r}) \phi_{j,1}({\bf r})+\phi^{\ast}_{j,2}({\bf r})
 \phi_{j,2}({\bf r}) \right] } }
\end{eqnarray}
where ``Im'' indicates the imaginary part, and the fact that $\sum_{j=1}^{N \over 2} \left[ \phi^{\ast}_{j,1}({\bf r}) \nabla \phi_{j,1}({\bf r})+\phi^{\ast}_{j,2}({\bf r})\nabla \phi_{j,2}({\bf r}) \right] $ is real (due to the fact that $\phi_{j,1}({\bf r})$ and $\phi_{j,2}({\bf r})$ are time-reversal partners) is used.

As is shown in the previous sections, optimal ${\bf A}^{\rm fic}_1$ and ${\bf A}^{\rm fic}_2$ are given by ${\bf A}^{\rm fic}_1={\bf A}^{\rm fic}_2={\bf A}^{\rm fic}=-{ \hbar \over {2e}} \nabla \chi$.
 In this case,  we have
\begin{eqnarray}
 {\bf A}^{\rm MB}={ e \over \hbar} {\bf A}^{\rm fic}=-{ 1 \over 2} \nabla \chi
 \label{AMB}
\end{eqnarray}
thus ${\bf A}^{\rm fic}$ is identified as ${\bf A}^{\rm MB}$ with factor ${ \hbar \over e}$ .
We may identify $\theta$ as $-\chi/2$.

The kinetic energy part of the Hamiltonian is given by
\begin{eqnarray}
K_0={ 1\over {2m}} \sum_{j=1}^{N} \left( {\hbar \over i} \nabla_{j} \right)^2
\label{a2}
\end{eqnarray}
where $m$ is the electron mass and $\nabla_{j} $ is the gradient operator with respect to the $j$th electron coordinate ${\bf r}_j$.

Using $\Psi$ and ${\bf A}^{\rm MB}$, we can construct a currentless wave function $\Psi_0$ for the current operator associated with $K_0$
\begin{eqnarray}
\Psi_0 ({\bf x}_1, \cdots, {\bf x}_{N},t)=\Psi ({\bf x}_1, \cdots, {\bf x}_{N},t)\exp\left(-{i } \sum_{j=1}^{N} \int_{0}^{{\bf r}_j} {\bf A}^{\rm MB} ({\bf r}',t) \cdot d{\bf r}' \right)
\label{wavef0}
\end{eqnarray}

Reversely, $\Psi ({\bf x}_1, \cdots, {\bf x}_{N},t)$ is expressed as
 \begin{eqnarray}
\Psi ({\bf x}_1, \cdots, {\bf x}_{N},t)=\Psi_0({\bf x}_1, \cdots, {\bf x}_{N},t)\exp\left(-{ i \over 2}\sum_{j=1}^{N} \chi ({\bf r}_j, t) \right)
\label{f0}
\end{eqnarray}
using the currentless wave function $\Psi_0$. 

Due to the spin-twisting the winding number of $\chi$ is non-zero, thus, a line of singularities exist within the loop around which non-zero winding number is obtained. The flux threaded through the line of singularities can be calculated using ${\bf A}^{\rm fic}$, and yields $\pi \mbox{ (mod 2 $\pi$)}$ ; thus, the line of singularities is the $\pi$-flux Dirac string. The form of the wave function in Eq.~(\ref{f0}) indicates that a collective mode
described by $\exp\left(-{ i \over 2}\sum_{j=1}^{N} \chi ({\bf r}_j, t) \right)$ that produces the persistent current exists.

In the present formalism, the superconducting state is the one with nontrivial ${\bf A}^{\rm fic}$. It plays dual roles; it is a Berry connection that enables the comparison of the phase of the wave function at different spatial points and gives rise to the macroscopic quantum interference effects; at the same time it is the collective mode $\nabla \chi$ of electrons with a long range order of the average momentum \cite{London1950}.

\section{Concluding Remarks}

In the present work, we have shown that the spin-twisting itinerant motion occurs for the conduction electrons of metals due to the Rashba spin-orbit interaction, and it generates ${\bf A}^{\rm fic}$. When the energy gap formation by electron pairing described by the BCS theory occurs, ${\bf A}^{\rm fic}$ is stabilized; then, the sum of ${\bf A}^{\rm fic}$ and ${\bf A}^{\rm em}$ form the effective gauge potential ${\bf A}^{\rm eff}$.

In our pervious work \cite{koizumi2019}, the state with the wave function Eq.~(\ref{f0}) is shown to be stabilized by the pairing interaction through the fluctuation of the number of particles participating in the collective mode $\nabla \chi$. 
Let us examine this point below.

In such a case, the ground state is given by
\begin{eqnarray}
|{\rm Gnd}({\bf r}; N)\rangle=\prod_{\bf k} \left( u_{\bf k}({\bf r}) + v_{\bf k}({\bf r})  c^{\dagger}_{{\bf k} \uparrow}c^{\dagger}_{-{\bf k} \downarrow} e^{- {i}\hat{\chi}({\bf r}) }\right) |{\rm Cnd} (N)\rangle
\label{newGnd}
\end{eqnarray}
where $N$ is the total number of particles, $e^{- {i}\hat{\chi}({\bf r}) }$ is the number changing operator that satisfies
\begin{eqnarray}
e^{- {i}\hat{\chi}({\bf r}) }|{\rm Cnd} (N)\rangle=e^{- {i}{\chi}({\bf r}) }|{\rm Cnd} (N-2)\rangle
\label{chiphase}
\end{eqnarray}
and $|{\rm Cnd}(N)\rangle$ is the state vector related to the wave function $\Psi_0$ in Eq.~(\ref{wavef0}) by
\begin{eqnarray}
\Psi_0({\bf x}_1, \cdots, {\bf x}_{N},t)=\langle {\bf x}_1, \cdots, {\bf x}_{N} |{\rm Cnd}(N)\rangle
\label{Cnd}
\end{eqnarray}

The number changing operator $e^{- {i}\hat{\chi}({\bf r}) }$ is obtained by noting that the conjugate momentum of $\chi$, $p_{\chi}$ is given by
\begin{eqnarray}
p_{\chi}={1 \over 2} \hbar \rho
\label{momentumchi}
\end{eqnarray}
where $\rho$ is the number density calculated with $\Psi_0$\cite{koizumi2019}.

Thus,
the canonical quantization condition, 
\begin{eqnarray}
[\hat{p}_{\chi}({\bf r}, t), \hat{\chi}({\bf r}', t)]=-i\hbar \delta ({\bf r}- {\bf r}')
\end{eqnarray}
where $\hat{p}_{\chi}$ and $\hat{\chi}$ are operators corresponding to ${p}_{\chi}$ and ${\chi}$ respectively, yields
\begin{eqnarray}
\left[{ {\hat{\rho({\bf r}, t)}} \over 2} , \hat{\chi}({\bf r}', t) \right]=-i \delta ({\bf r}- {\bf r}')
\label{commu0}
\end{eqnarray}
where $\hat{\rho}$ is the operator corresponding to $\rho$. 

From the above relation, $e^{-{i }\hat{\chi}({\bf r}) }$ is the number changing operator for the number of particles participating in the collective mode $\nabla \chi$ at ${\bf r}$ that satisfies Eq.~(\ref{chiphase})\cite{koizumi2019}.

The ground state $|{\rm Gnd}({\bf r}; N)\rangle$ actually corresponds to the BCS ground state with the phase factor $e^{- {i}{\chi}({\bf r}) }$
\begin{eqnarray}
|{\rm BCS} ({\bf r}; \chi) \rangle=\prod_{\bf k}\left(u_{\bf k}({\bf r})+v_{\bf k}({\bf r})c^{\dagger}_{{\bf k} \uparrow}c^{\dagger}_{-{\bf k} \downarrow}
e^{- {i}{\chi}({\bf r}) } \right)|{\rm vac} \rangle.
\label{BCSr}
\end{eqnarray}
in the new formalism\cite{koizumi2019}. 

As the similarity between $|{\rm Gnd}({\bf r}; N)\rangle$ and $|{\rm BCS} ({\bf r}; \chi) \rangle$ indicates, the mathematical structure is unaltered in the new formalism. However,  we can calculate superconducting properties with keeping the total particle number fixed in the new formalism.
 
There also exists an exception in relation to the ac Josephson effect.
The new formalism seems to be more in accordance with the observed ac Josephson effect.
 The recent re-derivation of the ac Josephson effect
  \cite{Koizumi2011,HKoizumi2015}
indicates that the boundary condition considered by Josephson and the one employed in the experiment are different. If the experimental boundary condition is employed and the care is taken for the gauge invariance, the observed Josephson relation actually indicates the charge on the particle is $q=-e$ not $q=-2e$. The new formalism explains the experimental ac Josephson effect with $q=-e$ since the role of the electron pairing 
is the stabilization of ${\bf A}^{\rm fic}$ and the supercurrent is the collective motion of electrons by $\nabla \chi$\cite{koizumi2019}.

Note that usually, the transfer of electron pairs is considered between the two superconductors in the Josephson junction using the second order perturbation theory by taking the usual electron transfer Hamiltonian as a perturbation; in this case, the supercurrent that flows without Bogoliubov excitations requires electron-pair tunneling.
However, the $q=-e$ electron transfer is possible if the two superconductors in the junction is in such a close contact that the Bogolibov quasiparticle excitations are absent during the electron transfer between them with simultaneous transferring of electrons between the superconductors and the external leads connected to them \cite{Koizumi2011,HKoizumi2015,koizumi2019}.
The present work suggests that ac Josephson effect actually occurs for the above close contact junction.
This point may be clarified if the re-investigation on the contact effect for the ac Josephson effect is performed. 
In this respect, it is noteworthy that un-paired electrons seem to be more abundant than the standard theory in  a Cooper pair box \cite{Martinis2004}.
 
 
%

\end{document}